\documentstyle[prd,aps,psfig]{revtex}
\def\beq{\begin{equation}}
\def\eeq{\end{equation}}
\def\bea{\begin{eqnarray}}
\def\eea{\end{eqnarray}}
\def\b {\beta}
\def\a {\alpha}
\def\s {\sigma}
\def\r {\rho} 
\def\O {\Omega}
\def\o {\omega}

\def\g {\gamma}

\def\laq{\raise 0.4ex\hbox{$<$}\kern -0.8em\lower 0.62
ex\hbox{$\sim$}}
\begin{document}
\draft

\twocolumn[\hsize\textwidth\columnwidth\hsize\csname 
@twocolumnfalse\endcsname

\title{ Kalb-Ramond axion production in anisotropic  
	string cosmologies}
\author{Ruth Durrer${}^1$ and Mairi Sakellariadou${}^2$}

\address{~\\  
${}^1$D\'epartement de Physique 
Th\'eorique, Universit\'e de Gen\`eve, 
\\
24 quai E.~Ansermet,  CH-1211 Gen\`eve 4, 
Switzerland.
\\~\\
${}^2$DARC, Observatoire de Paris, \\
UPR 176 CNRS, 92195 Meudon Cedex, France.}

\date{March 2000}
\maketitle

\begin{abstract}
We compute the  energy spectra 
for massless Kalb-Ramond axions in  four-dimensional anisotropic string 
cosmological models. We show that, when integrated over directions,
the four-dimensional anisotropic model leads to  infra-red
divergent spectra similar to the one found in the isotropic case.
\end{abstract}

\pacs{PACS number(s): 98.80.Cq }
\vspace{1cm}]

\section{Introduction}
The pre-big-bang (PBB) model of cosmology~\cite{PBB} inspired by the duality 
properties of string theory, is faced, on the 
phenomenological side, with the question whether or not it can
reproduce the amplitude and 
slope of the observed temperature anisotropy spectrum~\cite{cmb} 
and of large-scale density perturbations. 

Within the PBB model, the inflationary expansion due to the dilaton field in 
the low-energy effective action of string theory, leads to an
amplification of metric fluctuations  
as well as of quantum fluctuations of all the 
fields present in PBB cosmology. Such fields, which are not
part of the homogeneous background whose perturbations we 
study, are for example the gauge fields and the pseudo-scalar
partner of the dilaton field in the string theory effective 
action.
\par
At first, it was thought that the PBB scenario could not lead
to the observed scale-invariant Harrison-Zel'dovich spectrum 
of perturbations at large-scales.
First-order scalar and tensor metric perturbations were 
found to lead to primordial spectra that grow with 
frequency~\cite{stp}, with a normalization imposed by the 
string cut-off at the shortest amplified scales. These blue 
spectra have too little power at scales relevant for the 
observed anisotropies in the cosmic microwave background 
(CMB). In contrast, the axion energy spectra were found to 
be diverging at large scales, red spectra,
leading to very large CMB anisotropies,
in conflict with observations.
\par
These results already rule out four-dimensional isotropic PPB
cosmology. However, if one allows for internal contracting 
dimensions in addition to the three expanding ones,
the situation is different. The axion field can lead to a 
flat Harrison-Zel'dovich spectrum of fluctuations for an 
appropriate relative evolution of the external and the 
compactified internal dimensions~\cite{axion1,axion2}. Thus, it is 
possible that the amplification of quantum fluctuations of 
fields which are present  in the PBB scenario, can 
generate via the seed mechanism~\cite{seed} the observed 
anisotropy of the CMB radiation. 
\par
Considering an isotropic PBB model with extra dimensions, the amplification of 
electromagnetic vacuum fluctuations and of Kalb-Ramond axion 
vacuum fluctuations lead to interesting observational consequences within 
the context of primordial magnetic fields~\cite{mmg} and 
large-scale temperature 
anisotropies~\cite{rmmg}. In particular, massless axions as 
well as very light axions can exhibit a flat or slightly 
tilted blue spectrum which may reasonably fit the observational 
data~\cite{rmmg,afrg}. (Even though an acoustic peak at $\ell\sim 350$
is excluded by experiments published after Ref.~\cite{afrg} was completed,
it is possible to shift this peak to $\ell\sim 220$ by closing the
universe with a cosmological constant. More details about this model
can be found in Ref.~\cite{farg}.)
\par 
Recently it has been  suggested  that four-dimensional string
cosmology models which expand anisotropically can also lead 
to blue or flat energy spectra for axionic 
perturbations~\cite{aniso}. According to Ref.~\cite{aniso}, 
one can instead of assuming internal extra 
dimensions~\cite{rmmg}, consider an anisotropic 
four-dimensional background. This has become especially interesting
in view of new results which show that the pre-big-bang phase may
generically be homogeneous but anisotropic~\cite{kr}.

In  Ref.~\cite{aniso}, the axion spectrum is
only computed for the part of phase space where the longitudinal
component of the wave vector is sufficiently large. In this work we
correct the result of  Ref.~\cite{aniso} and complete the computation
to contain all directions in phase space. We then integrate the obtained
spectrum over directions and compare it with the result for the
isotropic PBB. We find that the anisotropic spectrum, when averaged
over directions agrees roughly with the isotropic
one. Therefore, anisotropic expansion during the pre-big-bang phase 
cannot solve the axion problem of four-dimensional string cosmology.

\section{Axion production in the pre-big-bang cosmological model}
Let us consider a four-dimensional spatially flat anisotropic
PBB cosmological model, with metric
\beq
 \left(g_{\mu\nu}\right)={\rm diag}[1,-a^2(t),-b^2(t),-b^2(t)]~;
\eeq
the internal compactified radii (if present) are assumed to 
be frozen. For simplicity, we assume two directions to expand with the
same scale factor $b$. Varying the low-energy string theory effective 
action (in the string frame)
\bea
S=&&-{1\over 2\lambda_s^2}\int d^4x\sqrt{-g}~e^{-\phi}
\nonumber\\
&&
\times\left[R+g^{\a\b}\partial_\a\phi\partial_\b\phi
-{1\over 12}H_{\mu\nu\a}H^{\mu\nu\a}\right]
\label{s}
\eea
($H^{\mu\nu\a}$ denotes the antisymmetric tensor field),
with respect to the metric and the dilaton field 
$\phi$, we obtain the dilaton driven vacuum solutions of
the tree-level evolution equations. As derived in 
Ref.~\cite{PBB}, these solutions read
\beq
a(\eta)=\left[-{\eta\over\eta_1}\right]^{\a\over 1-\a}~~,~~
b(\eta)=\left[-{\eta\over\eta_1}\right]^{\b\over 1-\a}~,
\label{ab}
\eeq
and
\beq
\phi(\eta)=\left({\a+2\b-1\over 1-\a}\right) \log\left
[-{\eta\over\eta_1}\right]~,
\label{dil}
\eeq
with $\a$ and $\b$ satisfying the Kasner condition
\beq 
\a^2+2\b^2=1~.
\eeq
Here $\eta$ denotes conformal time with respect to the scale factor
$a$. It is negative during the pre-big-bang era and $\eta=-\eta_1$ stands 
for the transition time from the dilaton driven pre-big-bang era to the 
radiation dominated post-big-bang era. To obtain the axion evolution 
equation, we vary the effective action,
Eq.~(\ref{s}),  with respect to the Kalb-Ramond axion field 
$\s$, given by
\beq
H^{\mu\nu\a} = e^\phi{\epsilon^{\mu\nu\a\rho}\over
\sqrt{-g}}\partial_\rho\s~.
\eeq
The evolution equation of the canonical field 
$\psi=e^{\phi/2} b \s$, in Fourier space, reads~\cite{aniso}
\beq
\psi_k'' +\left[k_L^2 +k_T^2{a^2\over b^2} - 
{{\cal P}''\over {\cal P}}\right]
\psi_k=0~~~,~~~{\cal P}=e^{\phi/2} b~,
\label{evax}
\eeq
where $k_L$ denotes the modulus of the comoving longitudinal 
momentum and $k_T=\sqrt{k_y^2+k_z^2}$ is the modulus of 
the transverse momentum. 
Equation (\ref{evax}) describes the generation of axionic 
modes, where the anisotropy of the spacetime
has been translated into an asymmetry between the
longitudinal and transverse momenta.

The choice $\a=\b=-1/\sqrt 3$ 
corresponds to the isotropic 
case, for which $a=b$ and the evolution of axionic fluctuations is
given by
\bea
&&\psi_k'' +\left[k^2  - {{\cal P}''\over {\cal P}}\right]
\psi_k=0~,\label{isoax}\\
\mbox{with}~~ &&\psi_k={\cal P}\s_k~~,~~
{\cal P}=e^{\phi/2} a \propto (-\eta)^p ~,\\
\mbox{so that } &&  {{\cal P}''\over {\cal P}} = {(\mu^2-1/4)\over
\eta^2}, \mbox{ with }~~ \mu^2=(p-{1\over 2})^2~.
\label{eviso}
\eea
The solution of Eq.~(\ref{isoax}), normalized to an initial vacuum 
fluctuation spectrum, can be written as
\beq
\psi_k=\eta^{1/2} H_\mu^{(2)}(|k\eta|) ~,~ \mu=
\left|p-{1\over 2}\right| ~,~\eta\le -\eta_1~,
\eeq
with $\mu=\sqrt 3$. $H_\mu^{(2)}$ denotes the Hankel function of
second kind (we adopt the conventions of Ref.~\cite{abra}). 
\par
Assuming that the dilaton driven era is followed by a radiation
dominated era, the density parameter of produced 
Kalb-Ramond axions per logarithmic frequency interval is~\cite{axion1}
\beq
\O_\s(\o,\eta)={\r(\o)\over \r_c}={1\over \r_c}{d\r_\s\over 
d\log \o}\simeq
g_1^2\O_\gamma(\eta)\left[{\o\over \o_1}\right]^{3-2\mu}~,
\eeq
where $\r(\o)$ denotes their spectral energy density 
 and $\r_c=3M_p^2H^2/(8\pi)$ stands for the
critical energy density. Note that
$\o_1 = k_1/a_1=1/(a_1|\eta_1|)$ represents the maximal amplified 
frequency, $g_1=H_1/M_p$ is the transition scale in units 
of the Planck mass, $H_1\simeq \o_1$ denotes the Hubble
 scale at which the universe becomes
radiation dominated. Hence  $\O_\gamma(\eta)=
(H_1/H)^2(a_1/a)^4$ is the radiation density parameter at a given time $\eta$.

Clearly a flat spectrum corresponds to $\mu=3/2$ and the value
$\mu=\sqrt{3}$ obtained in a four-dimensional isotropic pre-big-bang
model implies a red spectrum, leading to an unacceptable divergence at
low frequencies. 
\par
Let us now go back to the case of a four-dimensional anisotropic
background. We first study the evolution of  
axionic fluctuations and we then calculate 
the spectral energy density of the axionic inhomogeneities
($d\rho_\s/d\log\o$), as they re-enter the horizon during the
isotropic radiation dominated era, after being amplified 
during the anisotropic dilaton driven era. Inserting 
Eqs.~(\ref{ab}) and (\ref{dil}) into 
Eq.~(\ref{evax}), we obtain~\cite{aniso}
\beq
\psi_k'' +\left(k_L^2 +k_T^2\left[-{\eta\over\eta_1}
\right]^{\g}-{\mu^2-1/4\over \eta^2}\right)\psi_k=0~,
\label{evax2}
\eeq
where
\bea
\g &={2(\a-\b)\over 1-\a}~,~2\mu &
=|2p-1|~,~ \mbox{ where }\\
 p& ={\a+4\b-1\over 2(1-\a)}~, 2\mu &= 2-{4\b\over 1-\a}~.
\label{eq}
\eea
\par
If $\gamma<0$, the $k_T$-term as well as the $\eta^{-2}$-term go to zero for 
$\eta\rightarrow -\infty$; and initially the parentheses in Eq.~(\ref{evax2}) is
dominated by $k_L^2$ (except if $k_L\equiv 0$). If $k_T$ is not very
large, namely if
\beq
	k_T<k_L(k_1/k_L)^{-\g/2}~, \label{kTkL}
\eeq
the scale $k_L$ becomes super-horizon, {\em i.e.} the parentheses in
Eq.~(\ref{evax2}) is dominated  by the $1/\eta^2$-term, before
the $k_T$-term takes over. In this case, we may entirely neglect the
$k_T$-term in Eq.~(\ref{evax2}), which then  reduces to
Eq.~(\ref{isoax}) with $k$ replaced by $k_L$. Therefore, the spectrum
for these modes is flat for
$\mu=3/2$ which corresponds to 
\beq
	 \a=-7/9~~,~~~~ \b=-4/9 ~~\mbox{ and }~~~~ \g=-3/8~. 
\eeq
We also require the solution to expand,
{\em i.e.} $\a, \b < 0$. We first concentrate
mainly on these values of the Kasner exponents since they lead to a scale
invariant spectrum of fluctuations for directions with a sufficiently
large $k_L$-component, but we express our results in
terms of $\a$ and $\b$ so that they can then also be applied also to other
values of the Kasner indices. In the part of phase-space defined
by the inequality given in Eq.~(\ref{kTkL}), the energy density of the
produced axions has already been determined in Ref.~\cite{aniso}. 
Here we correct the result
of Ref.~\cite{aniso} and generalize it to the entire phase space.
\par
To solve Eq.~(\ref{evax2}), we distinguish among the following 
two cases:\\
{\bf (I)}  The modulus of the longitudinal momentum, $k_L$,  
always  dominates until $\eta^2<1/k_L^2$ at which point the
$1/\eta^2$ term comes to dominate. This is equivalent to the
condition given in Eq.~(\ref{kTkL}).\\
{\bf (II)} At some conformal time 
$\eta=\eta_T < -\eta_1$, the modulus of the transverse momentum, 
$k_T$, comes to dominate over $k_L$, but the mode  is still well within the 
horizon, {\em i.e.} $\sqrt{k_L^2 +k_T^2(-\eta_T/\eta_1)^\g}
  >\eta_T^{-2}$. Equation~(\ref{evax2}) implies
\beq
\eta_T=-\eta_1\left({k_L\over k_T}\right)^{2/\gamma}~.
\eeq
{\bf Case (I)~:} Let us first discuss this case which is also the one
studied in Ref.~\cite{aniso}. Here, the inequality given in
Eq.~(\ref{kTkL}) holds.
For low frequency modes, $\o\ll\o_1$
this is the case outside a very thin slice around the plane
$k_L=0$ if $\g<0$. In this situation we may entirely neglect the second term 
inside the parentheses of
Eq.~(\ref{evax2}) which yields a Bessel differential equation. Its 
solution during the pre-big-bang era , is simply
\bea
\psi_k^{\rm PBB}(k,\eta) = \sqrt{{|k_L\eta|}\over k_L}
~~&&H^{(2)}_{\mu}(k_L\eta)~,\nonumber\\
&&
\mbox {for~~~} \eta\le -\eta_1~,
\eea

After the transition to the radiation dominated era (RD), we assume
the dilaton to be frozen and the expansion to have become
isotropic. This implies ${\cal P}''=0$, $a/b=1$ and Eq.~(\ref{evax})
reduces to a simple harmonic equation with  general solution  
\bea
\psi_k^{\rm RD}(k,\eta) = {1\over\sqrt{ k}}\!
~~&&\left[c_+e^{-ik(\eta+\eta_1)}+c_-e^{ik(\eta+\eta_1)}
\right],\label{psiRD}\\
&&
\mbox {for~~~} \eta\ge -\eta_1~.
\eea
By matching the in-coming solution $\psi_k^{\rm PBB}$ to
the out-going one $\psi_k^{\rm RD}$, and by also matching 
their first derivatives, at the transition time 
$\eta =-\eta_1$, we obtain the frequency mixing 
coefficient $c_-(k)$:
\beq
c_- = {-1\over \sqrt{2\pi}} \sqrt{{1\over 
(k\eta_1)(k_L\eta_1)^{2\mu}}}~.
\label{c-}
\eeq
The coefficient $c_-$ determines the occupation numbers 
of produced axions. The spectral energy density of the
produced axions reads
\beq
\rho_L(\o,s) ={d\rho_\s\over d\log\o}\approx {\o^4\over
\pi^2}|c_-(\o)|^2~.
\label{sed}
\eeq
From Eqs.~(\ref{c-}),~ (\ref{sed}) we obtain with $\o=k/a$ for $\mu=3/2$
\beq
\rho_L(\o,s) \approx {1\over 2\pi^3} \o_1^4/s^{3}~,
\eeq
where ${s}=k_L/k$.
\par
Thus, if the longitudinal momentum $k_L$
dominates, the spectrum of produced Kalb-Ramond axions 
is flat, {\em i.e.} independent of $\o$, but anisotropic.
This result generically agrees with the finding of Ref.~\cite{aniso} (up to a
factor $1/s^2$, which we think is missing in Ref.~\cite{aniso}).
\par
{\bf Case (II)~:}
We now assume that the $k_T$-term comes to dominate before the
perturbation becomes superhorizon. As long as the perturbation is
sub-horizon,  we may approximate Eq.~(\ref{evax2}) by
\beq
\psi_k'' +\left(k_L^2 + k_T^2\left[-{\eta\over\eta_1}
\right]^{\g}\right)\psi_k=0~,
\label{evax3}
\eeq
An approximate solution to this equation is
\beq
\psi \simeq {\exp\left(\eta\sqrt{k_L^2 +q^2(-\eta/\eta_1)^{\g}k_T^2}\right)
	\over \sqrt{\pi/2}\left[k_L^2 +(-\eta/\eta_1)^{\g}k_T^2\right]^{1/4}}~,
	 \label{vacT}
\eeq
with $q=1/(1+\g/2)=(1-\a)/(1-\b)$.

In the regime considered, 
 $\eta\sqrt{k_L^2+q^2(-\eta/\eta_1)^{\g}k_T^2}\gg 1$, this solution
becomes exact, if either  $k_L$ or $k_T$ vanishes and it
is a good approximation if one of the two terms dominates. If the
 $k_L$-term and the $k_T$-term are of the same order, the relative error is
about $|\g/2|=3/16$. It is also clear that this represents the correctly
normalized incoming vacuum solution.

At conformal time $\eta=\eta_T$,
the transverse momentum $k_T$ comes to dominate
over the $k_L$- term in Eq.~(\ref{evax2}). 
At even later times, the $\eta^{-2}$-term will eventually dominate.
After $\eta_T$ Eq.~(\ref{evax2}) can be approximated by
\beq
\psi_k'' +\left(k_T^2\left[-{\eta\over\eta_1}
\right]^{\g}-{\mu^2-1/4\over \eta^2}\right)\psi_k=0~,
\label{evax4}
\eeq
with general solution~\cite{abra}
\bea
\psi_k(k_T,\eta) = && c_T^{(1)}\sqrt{|k_T\eta|} H_{\mu q}^{(1)}
\left({|k_T\eta|q} \left[{-\eta\over 
\eta_1}\right]^{\gamma/2}\right)\nonumber\\
&& -i c_T^{(2)}\sqrt{|k_T\eta|} H_{\mu q}^{(2)}
\left({|k_T\eta| q} \left[{-\eta\over \eta_1}
\right]^{\gamma/2}\right)~,
\label{sol}
\eea
where $q$ is as above, and $ H_{\mu q}^{(1)},
H_{\mu q}^{(2)}$ are Hankel functions of the 1st and 2nd kind of order $\mu q$.
For large $k_T|\eta|$ the second term just corresponds to the solution
(\ref{vacT}) in the limit where $k_L$ can be neglected. Therefore, by
matching the solutions we find
\bea
c_T^{(1)} &=&0\nonumber\\
c_T^{(2)} &=&{i\over\sqrt{k_T}}~,  \label{casII}
\eea
up to an irrelevant phase.

Next, we have to match the solution for the
field $\s$ of the pre-big-bang era to the solution for $\s$
during the radiation era at the transition time $\eta=-\eta_1$.

As we go from the pre- to the post-big-bang era, we assume the
universe to become isotropic and the dilaton field $\phi$ to become frozen.
Thus, here the matching of the in-coming to the 
out-going solution for $\s$, is not equivalent to matching  $\psi$. 
The relation between the the axion field $\s$ and the
canonical field $\psi$ at conformal time $\eta$ is
\bea
\s^{\rm RD}(\eta)&=&\left[{\eta\over \eta_1}\right]
^{-1}\psi^{\rm RD}(\eta)~,\\
\s^{\rm PBB}(\eta)&=&\left[-{\eta\over \eta_1}\right]
^{-\lambda}\psi^{\rm PBB}(\eta)~.
\eea
 The canonical field in Fourier space 
during RD is given in Eq.~(\ref{psiRD}).
Matching the solutions  and their first derivatives for $\s$, as we
pass  from PBB
to RD at time $\eta=-\eta_1$, we obtain for
$|k_T\eta_1|\ll 1$, the Bogoliubov coefficient
$c_-$  given by
\bea
|c_-|^2=&&\left[{\Gamma^2(\mu q)\over 4\pi^2}2^{2 \mu q}
\left({3\over 2}-{\mu  q}\right)^2\right]
\left({k_T\over k_L}\right)^{-2\mu q}\nonumber\\
&& \times  s^{-2\mu q}\left({\o\over\o_1}\right)
^{-1-2\mu q}~.
\eea
With Eq.~(\ref{sed}) we then obtain that the energy density of
the produced Kalb-Ramond axions, in the case where the
transverse momentum $k_T$ comes to dominate, {\em i.e} the inequality
given in Eq.~(\ref{kTkL}) is violated:
\bea
\rho_T(\o,s) =&&\left[{\Gamma^2(\mu q)\over 4\pi^2}2^{2 \mu q}
\left({3\over 2}-{\mu q}\right)^2\right]
{1\over\pi^2}\nonumber\\
&&\times \left({k\over k_T}\right)
^{2\mu q} \o_1^{1+2\mu q} \label{trans}
\o^{3-2\mu q}~.
\eea
Inserting the values $\mu=3/2, ~\a=-7/9 ~\b=-4/9$ which lead to a flat
spectrum in case~I one finds a somewhat blue spectrum in case~II, 
\beq
\rho_T(\o,s)
\propto \o^{-9/13}~.
\eeq
\par
Of course this case also gives a finite answer on the plane $k_L=0$
for which the result obtained under case~I diverges.

\section{Results and Conclusion}

In total we can summarize the calculated spectrum by
\beq
\rho(\o,s) \simeq {\o_1^4\over 2\pi^3}\left\{\begin{array}{l}
  s^{-2\mu} \left({\o\over \o_1}\right)^{3-2\mu }
	\mbox{ if }  k_T<k_L(k_1/k_L)^{-\g/2} \\
(1-s^2)^{-\mu q} \left({\o\over \o_1}\right)^{3-2\mu q}   ~\mbox{ else,}
\end{array}\right.
\eeq
where $s=k_L/k$~, $\mu = 1-2\b/(1-\a)$ and $q=(1-\a)/(1-\b)$.
For our prefered values, $\a=-7/9$ and $\b=-4/9$ which imply
$\mu=3/2$ and $q=16/13$, the above result reduces to
\beq
\O_\s(\o,s,\eta) \simeq g_1^2\O_\gamma(\eta)\left\{\begin{array}{l}
  s^{-3} 
	~~\mbox{ if }~~  k_T<k_L(k_1/k_L)^{-\g/2} \\
  (1\!-\! s^2)^{-{24\over 13}} \left({\o\over \o_1}\right)^{-9/13}
    \mbox{ else.}
\end{array}\right. \label{Omani}
\eeq
In the regime of phase space where the longitudinal mode of the momentum
is very small {\em i.e.} when the condition given in Eq.~(\ref{kTkL})
is  violated,
the spectrum of the produced Kalb-Ramond axions is not flat.
For a given value of $\o$, this is the case if $s$ is smaller than the
critical value $s_c$ which is well approximated by
\beq s_c(\o) \simeq \left({\o\over\o_1}\right)^{3/13}~, \mbox{ if }~~
	 \o \stackrel{<}{\sim} 0.1\o_1~,  \label{sc}
\eeq
a very small value for cosmologically interesting frequencies.
\par
In Figure~1 the the energy density $\rho(\o)$ is shown as a function
of $s$ for different values of $\o$.
\begin{figure}[ht]
\centerline{\psfig{figure=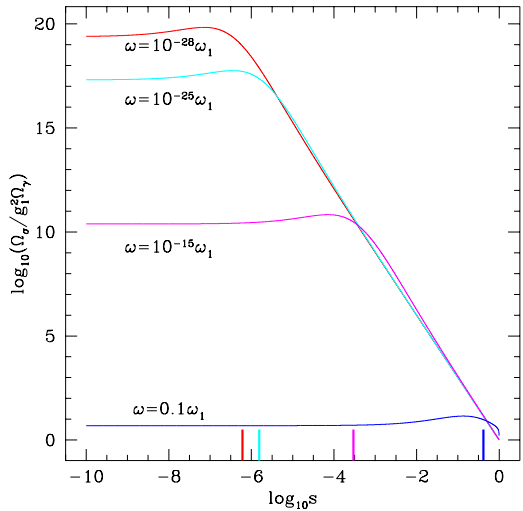,width=72mm}}
\caption{The density parameter of the produced Kalb-Ramond axions is
shown in units of $g_1^2\O_\g$ as a function
of $s$ for different values of $\o$. The value $\o_1\sim (0.01 -
0.1)M_{Pl}$ is the string scale. For this curve the results obtained
in Eq.~(\protect\ref{Omani}) have been interpolated logarithmically. The
little bars on the $\log(s)$-axis indicate the values of $s_c$ at
which the condition~(\protect\ref{kTkL}) becomes an equality. 
For values of $s$ smaller than 
$s_c(\o)$, the spectral density is approximated by $\rho_T(\o)$. }
\end{figure}
For $s$ fixed, if the modulus of the longitudinal
momentum dominates in Eq.~(\ref{evax2}),  more precisely if it satisfies
the condition $k_L>k_T(k_T/k_1)^{-\g/2/(1+\gamma/2)}$, the spectrum of the 
produced Kalb-Ramond axions is flat.

\par
To estimate the total energy density per logarithmic frequency interval 
we integrate the axion density $\O_\s(\o,s)$ over $s$.
For this we use
\beq
  d^3k=2\pi k_Tdk_L\wedge dk_T =4\pi k^2ds\wedge dk~,
\eeq
where we have used $dk_L=kds+sdk$ and
\[ dk_T={-s\over \sqrt{1-s^2}}kds + \sqrt{1-s^2}dk~. \]
Hence, we have
\bea
\lefteqn{\O_\s(\o,\eta) = \int\O_\s(\o,s,\eta)ds} \nonumber \\
 &\simeq&
{1\over \rho_c}\left[ \int_0^{s_c(\o)}\rho_T(\o,s)ds +  
	\int_{s_c(\o)}^1\rho_L(\o,s)ds\right] \label{gen} \\
 &\simeq& g_1^2\O_\gamma(\eta)\left[ s_c
	\left({\o_1\over \o}\right)^{9/13} + {0.5\over s_c^2}
	\right]. \label{Oo}
\eea
This spectrum is shown in Fig.~2.
\begin{figure}[ht]
\centerline{\psfig{figure=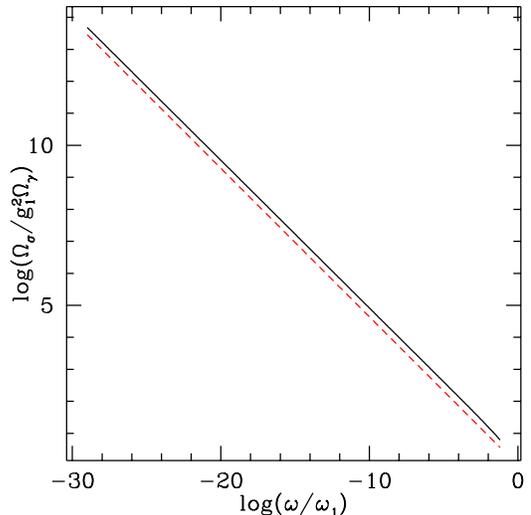,width=72mm}}
\caption{The energy density $\O_\s(\o)/g_1^2\O_\g$, integrated over
directions $s$ is shown as a function of $\o$ (solid line). Comparing
it with the isotropic result (dashed line) we conclude that,
the two spectral indices are the same and, within our accuracy, 
also the amplitudes are comparable.}
\end{figure}

Using $s_c \simeq (\o/\o_1)^{3/13}$, which is a good approximation as
long as  $\o\le 0.1\o_1$ it can also be seen directly from
Eq.~(\ref{Oo}) that the isotropic energy spectrum is nearly reproduced.
 The isotropic spectral index, $3-2\sqrt{3}\sim -0.464$ is actually replaced by
$-6/13\sim -0.463$.  Inserting reasonable values for the string scale, 
$0.01\le g_1<1$, we see that also in the
anisotropic case axions are over-produced in unacceptable amounts. Even
if the spectrum of the axions from wave vectors directed sufficiently
far from the plane $k_L=0$, is scale-invariant, the enhancement of the
spectrum in the vicinity of the plane  $k_L=0$ leads to a total 
contribution which agrees with the one obtained in the isotropic case.
Therefore,  the model is excluded 
(see Ref.~\cite{axion1}).

So far we have mainly considered the case $\a=-7/9$ and $\b=-4/9$, but
our results apply quite generically, as long as $\g<0$ and thus the
$k_L$-term dominates at sufficiently early times. But also if $\g>0$,
Eq.~(\ref{vacT}) is an approximate solution on sub-horizon scales. In
this situation, however the $k_T$-term dominates at sufficiently early
times and continues to do so until the perturbation becomes
super-horizon if the inequality given in Eq.~(\ref{kTkL}) is violated. For
$\g>0$ this is the case outside a narrow cylinder around the
$k_T=0$ axis. Therefore, the generic formula given in Eq.~(\ref{gen})
always applies, but $s_c \ll 1$, if $\g<0$ and  $s_c \simeq 1$, if $\g>0$.

For general values of $\a$ and $\b$ we obtain
\bea
\lefteqn{\O_\s(\o,\eta) = \int\O_\s(\o,s,\eta)ds} \nonumber \\
 &\simeq&
g_1^2\O_\gamma(\eta) \left[\left({\o\over\o_1}\right)^{3-2\mu q}
   \int_0^{s_c(\o)}(1-s^2)^{-\mu q}ds   \right. \nonumber \\ 
  && \left. ~~~~~~~ + ~~ \left({\o\over\o_1}\right)^{3-2\mu}\int_{s_c(\o)}^1
	s^{-2\mu}ds\right]~. \label{genint}
\eea
 The transition value of $s$ is given by
 \beq
  \sqrt{1-s_c^2}=s_c^{1+\g/2}\left({\o\over \o_1}\right)^{\g/2}~.
 \eeq
If $\g<0$ (i.e. $\a<\b$), the factor 
$\left({\o\over \o_1}\right)^{\g/2}$ is very 
large in most of phase space and hence $s_c\ll 1$. On the other hand, 
if $\g>0$ (i.e. $\a>\b$), the above factor is very small for the 
relevant frequencies, $\o\ll\o_1$ and $s_c \simeq 1$.
A reasonable approximation is
\bea
 s_c &\simeq \left({\o\over \o_1}\right)^{q-1} & \mbox{ if } ~~~\g<0 \\
 1-s_c^2 &\simeq \left({\o\over \o_1}\right)^{2/q-2} &  
	\mbox{ if } ~~~\g > 0~,
\eea
where we have used  the relation $ q=1/(1+\g/2)$.
Inserting these results in Eq.~(\ref{genint}), the integrals can be 
approximated by
\beq
\O_\s(\o,\eta) \sim g_1^2\O_\g(\eta)\left({\o\over \o_1}\right)^n~, 
	~~~\mbox{ where }
\eeq
\bea
n &=  2+q-2\mu q  & = {1+\a+2\b\over 1-\b}  ~\mbox{ if }  \a<\b~, \\
n &= 1+2/q-2\mu   & = {1+\a+2\b\over 1-\a} ~\mbox{ if }  \a > \b~. 
\eea
Clearly, since $\a^2+2\b^2=1$ and $\a,\b\le 0$ it is $\a+2\b\le -1$.
This shows that the spectrum is never blue and becomes scale invariant only in
the degenerate case with two static dimensions, $\b=0$. This is also 
shown in Fig.~3, where the above approximation for the 
spectral index  plotted as a function of $\a$: the spectrum
always remains red with a spectral index relatively close to the
isotropic value, $n_{\rm iso}=3-2\sqrt{3}\sim -0.46$, except in the extremal 
case, when two dimensions are frozen and $\a=-1$. 

If one relaxes the condition that both $a$ and $b$ be expanding and
just asks for volume expansion, $\a +2\beta<0$, there is another pair
of values for the Kasner indices leading to a flat spectrum, namely
$\a= -1/3$ and $\beta=-2/3$. However, if we want expansion in all
three dimensions the spectrum is always red.

\begin{figure}[ht]
\centerline{\psfig{figure=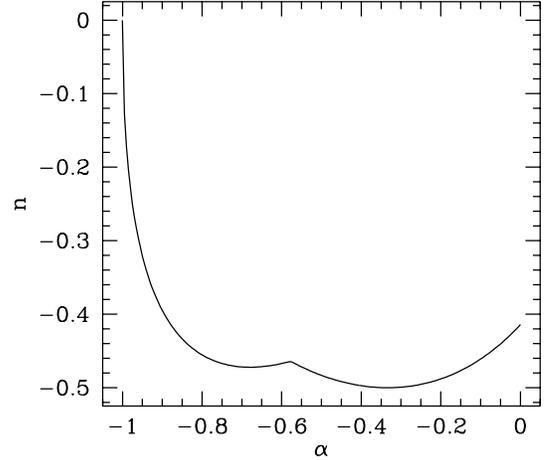,width=72mm}}
\caption{The  spectral index $n$ is shown as a function of the 
exponent $\a$ of the expansion law. For $\a = -1/\sqrt{3}$, 
the isotropic case, our approximation is not very good since
there $\a=\b$. This is reflected in the unphysical kink at this value
of $\a$. Clearly, the resulting spectrum is always
 red ($n<0$), with $ -0.4 > n\ > -0.5$ except close to  the degenerate 
case $\a \rightarrow -1$.}
\end{figure}

To summarize, we find that
anisotropic expansion has very little influence on the overall axion
production and cannot cure the axion problem of
four-dimensional pre-big-bang models. Only by allowing for extra 
dimensions one can escape this conclusion and obtain a scale invariant 
spectrum of axions as described in Refs.~\cite{rmmg,afrg}. 
A 'realistic' string cosmology 
with a Kalb-Ramond axion can therefore be realized only in models with
extra dimensions.

\acknowledgements
It is a pleasure to thank A. Buonanno, T. Damour, K.~Kunze, G. Veneziano and
A. Vilenkin for useful discussions. This work is supported by the
Swiss National Science Foundation.

\end{document}